\documentclass[11pt,twoside]{article}

\usepackage{asp2006}
\usepackage{fancyhdr,graphicx,caption,rotating,multirow,lscape}
\DeclareGraphicsExtensions{.eps,.eps.gz,.ps,.jpg,.tiff}

\begin{document}

\title{Searching for Variables in One of the WHAT Fields}  


\author{A. Shporer$^1$, T. Mazeh$^1$, A. Moran$^1$, G. Bakos$^2$ \& G. Kovacs$^3$}
\affil{$^1$Wise Observatory, Tel Aviv University\\
       $^2$Hubble Fellow, Harvard-Smithsonian Center for Astrophysics\\
       $^3$Konkoly Observatory, Hungarian Academy of Sciences}

\begin{abstract} 
We present preliminary results on a single field observed by WHAT,
a small-aperture short focal length automated telescope with 
an $8.2\deg \times 8.2\deg$ deg field of view, located at the Wise Observatory.
The system is similar to the members of HATNet 
(http://cfa-www.harvard.edu/$\sim$gbakos/HAT/) and is aimed at searching for 
transiting extrasolar planets and variable objects. 
With 5 min integration time, the telescope 
achieved a precision of a few mmag for the brightest objects. 
We detect variables with amplitudes less than 0.01 mag.
All 152 periodic variables are presented at 
http://wise-obs.tau.ac.il/$\sim$amit/236/.
\end{abstract}



\textbf{WHAT} is the \textbf{W}ise-observatory \textbf{H}ungarian-made 
\textbf{A}utomated \textbf{T}elescope, located at 
the Wise Observatory in Mizpe Ramon, Israel. It is operated under a 
collaboration between the Wise Observatory of the Tel-Aviv University and 
Konkoly Observatory of the Hungarian Academy of Sciences \citep{Shporer06}. 
Like all other HAT telescopes \citep{Bakos02, Bakos04}, WHAT is a 
combination of a fully automated telescope mount, a clamshell dome, 
2K $\times$ 2K CCD, 200mm $f/1.8$ Telephoto lens, RealTime Linux PC and a 
software environment, ``ProMount''. WHAT monitors the intensity of stars in 
the magnitude range of 
$V$ = 9 -- 15 in large fields of views, $8.2\deg \times 8.2\deg$ each, 
searching for transiting planets and variable stars.

Preliminary results of variability analysis for field \#236 
are presented here. 
Located at R.A. 15:28 and Dec. +30:00, this field was 
observed 3892 times in a period of 162 days between Feb. 20 and Aug. 1, 2004. 
Observations were carried out through a single $I$-band filter and 
PSF-broadening mode, whereby the telescope is stepped on a prescribed 
pattern to achieve a better sampled PSF. Exposure time was 300 seconds.

\begin{figure}[!ht]
\centering
\includegraphics[angle=0,width=8cm]{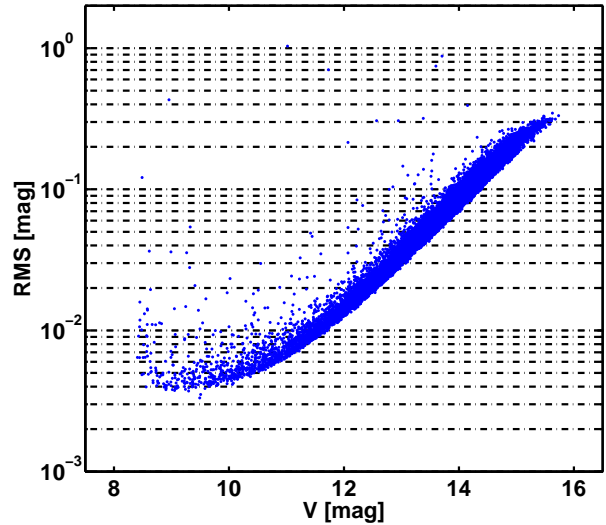}
\caption{Photometric results for WHAT field 236. Light curves RMS is plotted 
against approximate mean $V$ magnitude.\label{rms_mean}}
\end{figure}

Light curves of 13360 field stars, with over 2500 measurements each, 
were extracted using aperture photometry. Systematics were removed using 
the SysRem algorithm of \cite{Tamuz05}. RMS versus 
approximate $V$ mean magnitude for all objects are presented in 
Fig.~\ref{rms_mean}. The resulting scatter goes down to 4.5 milli-magnitude
for the brightest objects. For 1036 objects brighter than $V \approx 11.5$ mag 
the scatter is below 1\% and for 2324 objects brighter than $V \approx 12.4$ 
mag the scatter is below 2\%.

We\,\, detected\,\, 152\,\, periodic\,\, variables\,\, by\,\, using\, Analysis\,\, of\, Variance
\citep{Czerny89} and Lomb-Scargle power spectrum
\citep{Scargle82}. From these variables 133 are new ones. Periods
range from 0.09 days to 31 days, some of them with peak-to-peak
amplitude less than 0.01 mag. For the detection of periodic
variables, we also used the TFA algorithm of \cite{Kovacs05}. 
In addition, by using the Alarm statistics of \cite{Tamuz06}, we 
have identified 14 non-periodic variables -- from which 10 are new discoveries.

\begin{figure}[!ht]
\centering
\includegraphics[angle=0,width=11cm]{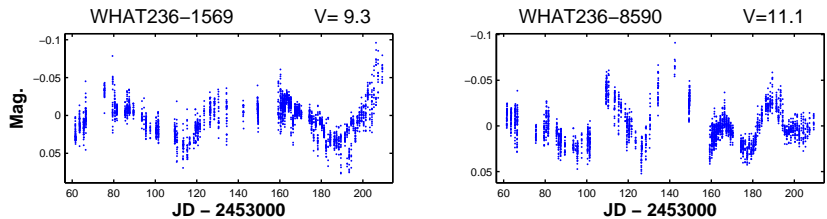}
\caption{Two new non-periodic variables detected by WHAT. 
Headers are, from left to right: Object internal ID
and approximate mean $V$ magnitude.
\label{nonper}}
\end{figure}

All variables are presented at: http://wise-obs.tau.ac.il/$\sim$amit/236/. A 
few new variables are presented below. Fig.~\ref{eb} presents four new 
eclipsing binaries, Fig.~\ref{puls} presents four new pulsating variables
and Fig.~\ref{nonper} presents two new non-periodic variables. 
All light curves are post-processed by SysRem (Tamuz et al. 2005).

\begin{figure}[!ht]
\centering
\includegraphics[angle=0,width=11cm]{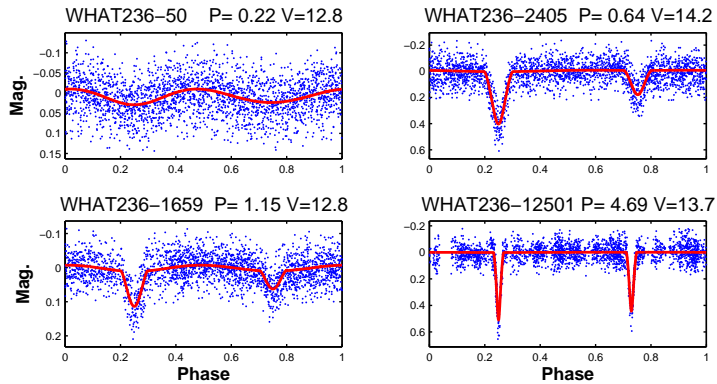}
\caption{Four new eclipsing binaries detected by WHAT. 
Continuous lines show the model fits by EBAS (Tamuz et al. 2006).
Headers are, from left to right: Object internal ID,
period in days and approximate mean $V$ magnitude.
\label{eb}}
\end{figure}

\begin{figure}[!ht]
\centering
\includegraphics[angle=0,width=11cm]{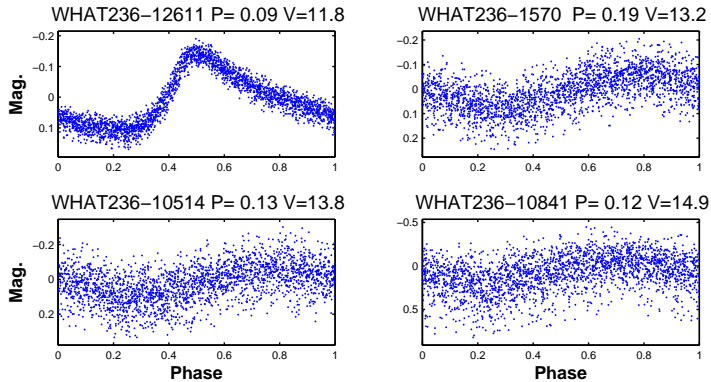}
\caption{Four new periodic pulsating variables detected by WHAT. 
Headers are as in Fig.~\ref{eb}.\label{puls}} 
\end{figure}

We are currently engaged with finalizing our results and processing data 
from additional fields. Further description of the WHAT project can be 
found at the WHAT homepage: http://wise-obs.tau.ac.il/$\sim$what.



\acknowledgements 

The WHAT project has been made possible by grants
from the Sackler Institute of Astronomy and from the Hungarian
Scientific Research Fund (OTKA M-041922, K-60750). Work of
G.~B. was supported by NASA through Hubble Fellowship grant
HF-01170.01 awarded by the STScI, which is operated by the AURA,
Inc., for NASA, under contract NAS 5-26555. This research was
supported by a Grant from the G.I.F., the German-Israeli Foundation 
for Scientific Research and Development.


\end{document}